\documentclass{aastex631}

\usepackage{natbib}

\begin{document}

\title{Acceleration of Solar Eruptions via Enhanced Torus Instability Driven by Small-Scale Flux Emergence}

\author[0000-0001-5121-5122]{Satoshi Inoue}
\affiliation{Center for Solar-Terrestrial Research, New Jersey Institute of Technology, Newark, 07102-1982, NJ, USA}

\author[0000-0002-4675-4460]{Takahiro Miyoshi}
\affiliation{Graduate School of Advanced Science and Engineering, Hiroshima University, Higashi-Hiroshima, 739-8526, Japan}

\author[0000-0001-9046-6688]{Keiji Hayashi}
\affiliation{Center for Solar-Terrestrial Research, New Jersey Institute of Technology, Newark, 07102-1982, NJ, USA}

\author{Huu Minh Triet Nguyen}
\affiliation{Center for Solar-Terrestrial Research, New Jersey Institute of Technology, Newark, 07102-1982, NJ, USA}

\author[0000-0002-8179-3625]{Ju Jing}
\affiliation{Center for Solar-Terrestrial Research, New Jersey Institute of Technology, Newark, 07102-1982, NJ, USA}

\author[0000-0003-2427-6047]{Wenda Cao}
\affiliation{Center for Solar-Terrestrial Research, New Jersey Institute of Technology, Newark, 07102-1982, NJ, USA}
\affiliation{Big Bear Solar Observatory, New Jersey Institute of Technology, 40386 North Shore Lane, Big Bear City, CA 92314, USA}

\author[0000-0002-5233-565X]{Haimin Wang}
\affiliation{Center for Solar-Terrestrial Research, New Jersey Institute of Technology, Newark, 07102-1982, NJ, USA}



\begin{abstract}
Despite decades of research, the fundamental processes involved in the initiation and acceleration of solar eruptions remain not fully understood, making them long-standing and challenging problems in solar physics. Recent high-resolution observations by the Goode Solar Telescope have revealed small-scale magnetic flux emergence in localized regions of solar active areas prior to eruptions. Although much smaller in size than the entire active region, these emerging fluxes reached strengths of up to 2000 G. To investigate their impact, we performed data-constrained magnetohydrodynamic (MHD) simulations. We find that while the small-scale emerging flux does not significantly alter the pre-eruption evolution, it dramatically accelerates the eruption during the main phase by enhancing the growth of torus instability, which emerges in the nonlinear stage. This enhancement occurs independently of the decay index profile. Our analysis indicates that even subtle differences in the pre-eruption evolution can strongly influence the subsequent dynamics, suggesting that small-scale emerging flux can play a critical role in accelerating solar eruptions.

\end{abstract}

\keywords{Solar flares(1496) --- Magnetohydrodynamics(1964) --- Solar active region magnetic fields(1975) --- Magnetohydrodynamical Simulations(1966)}

% ============================================================
\section{Introduction} \label{sec:intro}
% ============================================================
Solar eruptions, including solar flares and coronal mass ejections, are among the most energetic phenomena in the solar system. The mechanisms that trigger and drive these eruptions remain longstanding problems in solar physics \citep{Shibata2011}. Recent high-resolution observations have revealed clear magnetic disturbances preceding flares \citep{Kubo2007, Bamba2018}. Notably, the Goode Solar Telescope (GST; \citealt{Cao2012, Goode2010}) at Big Bear Solar Observatory (BBSO) observed an M6.5 flare in active region (AR) 12371 on 2015 June 22 \citep{Jing2016, Kang2019, Liu.Nian2022}. Figure~\ref{fig:f1}(a) shows AR 12371, as observed by the Helioseismic and Magnetic Imager (HMI; \citealt{Scherrer2012}) onboard the Solar Dynamics Observatory (SDO; \citealt{Pesnell2012}). This AR consisted of several sunspot groups, including a $\delta$-spot on the eastern side. The M6.5 flare occurred along the polarity inversion line (PIL) near this eastern sunspot. The inset in Figure~\ref{fig:f1}(a) shows a zoomed-in view of this region, observed by GST, corresponding to the white square in the main panel. A key observation was the detection of a small-scale emerging flux at the PIL, indicated by the green circle, just prior to the flare \citep{Wang2017}. Although this flux was much smaller in scale than the main sunspots, it exhibited opposite polarity and reached a peak strength of ~2000 G. While it was suggested that this emerging flux played a role in the eruption, the specific mechanism remains unclear.

In this Letter, we investigate the role of small-scale emerging flux in solar eruptions by performing data-constrained MHD simulations \citep{Inoue2016} of AR 12371, incorporating a realistic magnetic environment. Unlike previous studies that primarily focused on the triggering of eruptions \citep{Chen2000, Kusano2012, Muhamad2017, Jing2021, Torok2024}, our work emphasizes the influence of emerging flux during the dynamic, pre-eruption phase. The remainder of this Letter is organized as follows. Section 2 describes the observations and numerical methods. Section 3 presents the results, and Section 4 discusses and summarizes the key findings and their implications.

% ==================================================
\section{Observations and Numerical Methods} \label{sec:obs_Num}
% ===================================================
% -----------------------------------------------------------
\subsection{Observational data from the BBSO}
The high-resolution photospheric magnetogram shown in Figure 1(a) was obtained using the Near-infrared imaging spectropolarimeter (NIRIS \citealt{Cao2012}) of the Goode Solar Telescope (GST \citealt{Cao2010, Cao2012}). The center of the flaring sunspot group is about (6 degree, 13 degree) at 18:23 UT, corresponding to (90", 180") in heliocentric-Cartesian coordinates. Since it's very close to the disk center, the projection effect is negligible. The GST is a 1.6 m off-axis telescope that enables near-diffraction-limited observations of the sun with the aid of a high-order adaptive optics system. NIRIS is equipped with an advanced HAWAII-2 infrared detector and an improved dual Fabry-Perot interferometers system. The spectropolarimetric data from NIRIS were processed using the NIRIS data processing pipeline (\citealt{Anh2016}), which includes dark-field and flat-field corrections, instrumental cross-talk calibrations, and Milne-Eddington Stokes inversion. On 22 June 2015, the spatial resolution of the NIRIS magnetogram was $\sim 170~km$ at the 1.56 $\mu$m Fe I line with a bandpass of 10 pm, which was close to the GST diffraction limit. 

\subsection{Observational data from the SDO}
 We used the photospheric magnetic field obtained by HMI onboard SDO
 as the bottom boundary conditions for the NLFFF extrapolation and the data-driven MHD simulation. Specifically, we used the HMI photospheric magnetogram \verb#hmi.B_720s# data series taken between 16:36 UT and 16:48 UT on 22 June 2015, 1 h before the flare, to produce the magnetic field boundary conditions. Our magnetic field boundaries had a field of view of 480'' $\times$ 480'', centered at (150'', 175'') at 16:36 UT. We then transformed these data into a local Cartesian coordinate system using the same cylindrical equal area (CEA) projection that is used to produce the standard space-weather HMI AR patch (SHARP) format (\citealt{Bobra2014}).  We did not use the \verb#hmi.sharp_cea_720s# data provided by HMI, but produced our own CEA data, because the \verb#hmi.sharp_cea_720s# data for this active region are not sufficiently large in the Y direction for the simulation. The final bottom boundary of the magnetic field consists of 960 $\times$ 960 uniform grid points corresponding to $\sim$ 348 $\times$ 348 Mm$^2$.  

\subsection{Data-constrained MHD Simulation}
\subsubsection{Equations and Parameters}
Data-constrained MHD simulations employ the following zero-beta MHD equations (\citealt{Inoue2014,Inoue2018}):
  
% --------------------------------------------------------------------------
% Mass Equation
% --------------------------------------------------------------------------
  \begin{equation}
   \rho=|{\bf B}|,
   \label{eq_mass1}
  \end{equation}
% ------------------------------------------------------------------------
% Equation of Motion
% ------------------------------------------------------------------------
  \begin{equation}
  \frac{\partial {\bf v}}{\partial t} 
                        = - ({\bf v}\cdot {\bf \nabla}){\bf v}
                          + \frac{1}{\rho} {\bf J} \times {\bf B}
                          + \nu{\bf \nabla}^{2}{\bf v},
  \label{eq_motion}
  \end{equation}

% ------------------------------------------------------------------------
% Induction equation
% ------------------------------------------------------------------------
  \begin{equation}
  \frac{\partial {\bf B}}{\partial t} 
                        =  {\bf \nabla}\times({\bf v}\times{\bf B})
                        +  \eta {\bf \nabla}^2 {\bf B}
                        -  {\bf \nabla}\phi, 
  \label{induc_eq}
  \end{equation}

% ------------------------------------------------------------------------
% Ampere's low
% ------------------------------------------------------------------------
  \begin{equation}
  {\bf J} = {\bf \nabla}\times{\bf B},
  \end{equation}
  
% ------------------------------------------------------------------------
% Dedner 
% ------------------------------------------------------------------------
  \begin{equation}
  \frac{\partial \phi}{\partial t} + c^2_{h}{\bf \nabla}\cdot{\bf B} 
    = -\frac{c^2_{h}}{c^2_{p}}\phi,
  \label{div_eq}
  \end{equation}

where ${\bf B}$ is the magnetic flux density, ${\bf v}$ is the velocity, ${\bf J}$ is the electric current density, $\rho$ is the plasma density, and $\phi$ is a convenient scalar potential for reducing the errors in ${\bf \nabla}\cdot {\bf B}$ (\citealt{Dedner2002}). Note that for simplicity and to focus solely on the velocity$-$magnetic field interaction, we update the plasma density using Equation (\ref{eq_mass1})(\citealt{Amari1996,Inoue2014}). The length, magnetic field, density, velocity, and time are normalized by $L^{*}=3.48\times 10^{8}m$ , $B^{*}=0.28T$, $\rho^{*}=6.15\times 10^{-8}kg/m^3$ , $V_{\rm A}^{*}\equiv B^{*}/(\mu_{0}\rho^{*})^{1/2}=1.0\times 10^{6}m/s$, where $\mu_0$ is the magnetic permeability in free space, and $\tau_{\rm A}^{*}\equiv L^{*}/V_{\rm A}^{*}=348s$. The second-order central differential method was used to approximate the space derivative and the Runge-Kutta-Gill method was applied for time integration. The coefficients $c_h^2$ and $c_p^2$ in Eq.(\ref{div_eq}) are set with values 0.04 and 0.1, respectively. The coefficients $\nu$ and $\eta$ correspond to the viscosity and resistivity, respectively, given as $2.5\times 10^{-4}$ and $1.0\times 10^{-5}$. 
A numerical box with dimensions 1.0 $\times$ 1.0 $\times$ 1.0 in the non-dimensional scale is divided into 320 $\times$ 320 $\times$ 320 grid points. This was preceded by  $3 \times 3$ binning of the data points of the boundary data. 

% ==============================================================
\subsubsection{Boundary and Initial Conditions}
% ==============================================================
Regarding the boundary conditions outside the flux emergence region, the normal component of the magnetic field at the boundary was fixed in time, while the horizontal components evolved according to the induction equation. The velocity was set to zero, and the normal derivative of $\phi$ was prescribed. As a result, the temporal evolution of $B_x$ and $B_y$ at the bottom were according to
\begin{equation}
\frac{\partial B_x}{\partial t}=\frac{\partial E_y}{\partial z}+\eta\frac{\partial^2 B_x}{\partial z^2}
-\frac{\partial \phi}{\partial x},
\label{bot_bx}
\end{equation}
\begin{equation}
\frac{\partial B_y}{\partial t}=-\frac{\partial E_x}{\partial z} +\eta\frac{\partial^2 B_y}{\partial z^2}
-\frac{\partial \phi}{\partial y},
\label{bot_by}
\end{equation}
 where ${\bf E}=-{\bf v}\times{\bf B}$. Note that, $B_x$ and $B_y$ do not evolve in accordance with the observations. 
 
 We conducted five data-constrained MHD simulations: Runs R and R1, and Runs 1$-$3. Small-scale emerging flux was introduced in Runs 1$-$3, but not in Runs R and R1. See table-1 for details. All runs used an initial magnetic field condition generated from a data-driven MHD simulation (\citealt{Hayashi2018, Inoue2023b}), in which the NLFFF extrapolated at 16:36 UT (Figure \ref{fig:f2}(a)) was evolved for 0.375 Alfv\'en time to introduce a slight deviation from equilibrium. The detailed procedure is described in \citet{Inoue2023b}. A comparison between the NLFFF and the updated magnetic field is presented in Section \ref{sec:results}.
 
% ---------------------------------------------------------------
\subsection{Model of Small-scale Emerging Flux}
\begin{figure}[ht!]
\plotone{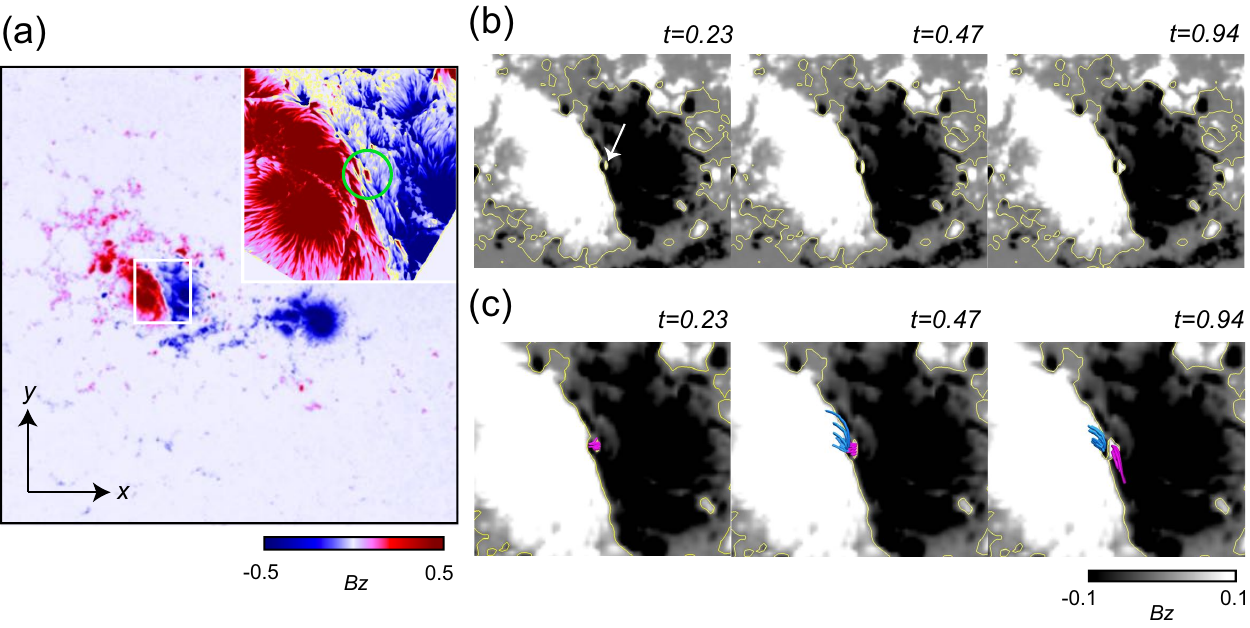}
\caption{(a) $B_z$ distribution of AR 12371 at 16:36 UT on 2015 June 22, one hour before the M6.5 flare observed in the eastern sunspots. Red and blue indicate positive and negative polarities, respectively. The simulation domain corresponds to this field of view. The inset shows a zoomed-in $B_z$ map of the eastern sunspots, which is enclosed by white square, at 17:35 UT on 2015 June 22, observed with the near-infrared imaging spectropolarimeter at BBSO. The yellow line marks the polarity inversion line (PIL), and the green circle highlights the small-scale emerging flux. (b) $B_z$ distribution on the eastern sunspot at 16:36 UT, along with the temporal evolution of the small-scale emerging flux, marked by white arrow. White and black indicate positive and negative polarities, respectively; the yellow line traces the PIL. (c) Magnetic field lines, which are traced from the small-scale emerging flux. Purple and light blue lines originate from the positive and negative polarities of the small-scale emerging flux, respectively. \label{fig:f1}}
\end{figure}

% ---------------------------------------------------------------
To match the GST observations, small-scale emerging flux was inserted from the subsurface at $t=0$. The flux emergence was modeled in the local area of the AR (See Figure \ref{fig:f1}(b)), and its evolution followed our previous work \citep{Jing2021}. A semi-spherical flux structure \citep{Kusano2012, Muhamad2017, Jing2021} was placed just below the photosphere, with its center located at $(x_e,y_e)=(0.355,0.51)$ and began to rise at $t=0$. During its ascent, the surrounding field evolved under data-constrained conditions. The magnetic field strength of the emerging flux $|B_{emf}|$ was treated as a parameter, with values of $0.75$, $0.5$, and $0.25$ assigned in Runs 1, 2, and 3, respectively. The polarity inversion line (PIL) of the emerging flux is oriented along the y-direction, with its polarity opposite to that of the eastern sunspot, where we set $\Phi = 0$, following the definition in \citet{Jing2021}. It is nearly parallel to the main PIL of the eastern sunspot. Since the magnetic field lines in the prescribed emerging flux are oriented perpendicular to the PIL, no twist is introduced in the emerging flux. The semi-sphere had a radius $r_c=5.0\times10^{-3}$, and an electric field $-{\bf V}_{emf}\times{\bf B}_{emf}$ was imposed on its cross section to drive upward motion with $V_{emf}=5.0\times 10^{-3}$, so that emergence ended at $t=r_c/V_{emf}=1.0$. After $t=1$, the emerging flux ceased. Before full emergence ($t \le 1$), the boundary magnetic field within the emergence region was fixed to the prescribed structure. After $t > 1$, all bottom boundary fields evolved according to the data-constrained simulation: the horizontal field evolved in the induction equation, while the normal component remained fixed. Figure \ref{fig:f1}(b) shows the field liens plotted from the small-scale emerging flux. Note that closed loops are initially prescribed in the emerging flux, and the subsequent evolution of the field lines is the result of the MHD simulation.

The unsigned magnetic flux of the emerging flux used in this study ranged from $1.73 \sim 5.2 \times 10^{-5}\Phi^*$, where $\Phi^*=B^*L^{*2}\approx3.4\times 10^{16}$Wb. A previous study by \cite{Wang2017} measured an increase in magnetic flux of approximately $4.7 \times 10^{-6}\Phi^*$. Therefore, the increase in magnetic flux in our simulations is still larger than that observed in AR 12371.

% ============================================
\subsection{Analysis of the Magnetic Field}
% ============================================
% --------- Twist -----------------------------------
 \begin{itemize}
 \item {\bf Twist value}: To detect twisted field lines, we calculated the magnetic twist number, $T_w$, for each field line using the following equation(\citealt{Berger2006}),
 \begin{displaymath}
     T_w =\frac{1}{4\pi} \int \frac{{\bf \nabla}\times {\bf B}\cdot {\bf B}}{B^2}dl,
 \end{displaymath}
where $dl$ denotes a line element. Note that $T_w$ measures the number of turns of two infinitesimally close field lines, which is distinct from the number of turns of the field lines around the magnetic axis of the MFR (\citealt{Threlfall2018}). 
Since negative helicity dominates in this active region, we focus on the negative twist $T_{wn}$ in this study. To avoid the calculation on the weak magnetic field area, the twist was calculated on the area where $|B_z|$ is larger than 0.01. \\

% ------ Decay Index ----------------------------------
\item {\bf Decay Index}: To examine the torus instability proxy(\citealt{Kliem2006}), we computed the decay index $n(x,y,z)$ in the 3D volume, 
\begin{displaymath}
n(x,y,z) = - \frac{z}{|{\bf B}_{ex}(x,y,z)|}\frac{\partial |{\bf B}_{ex}(x,y,z)|}{\partial z},
\end{displaymath}
where ${\bf B}_{ex}$ is the horizontal component of the external field. The decay index quantifies how rapidly the horizontal field decays with height and indicates torus instability when $n(x,y,z)\approx 1.5$ is reached (\citealt{Kliem2006}). Here ${\bf B}_{ex}$ was assumed to be a potential field (\citealt{Fan2010, Aulanier2010}). Note that this potential field differs from the actual field surrounding the MFR, which is non-potential even initially and becomes more so as the MFR evolves. Thus, the $n$ value provides only an approximate instability criterion.\\

% ----- Twist Flux -------------------------------------
\item {\bf Magnetic Flux of MFR}: The magnetic flux, $\Psi_M$, of the MFR is calculated using the following equation, 
\begin{displaymath}
\Psi_M = \int_{|T_w{_n}| \ge 1.0} |B_z| dS,
\end{displaymath}
where $B_z$ is the normal magnetic field component of the bottom surface and $dS$ denotes the surface element. First, we calculated the magnetic twist $T_{wn}$ by tracing the magnetic field lines from one footpoint to the other, and the twist value was plotted at each footpoint of the field lines. The magnetic flux satisfying $|T_{wn}| \ge 1.0$ is then calculated. We will discuss the threshold, $|T_{wn}| \ge 1.0$, later. \\

% ---- Current Intensity > Tw>1.0 ---------------------------
\item {\bf Toroidal current perpendicular to the MFR cross-section}: We calculate the electric current intensity at a specific area of the bottom surface which satisfies $|T_{wn}| \ge 1.0$. The toroidal current, $I_M$, flowing inside the MFR 
\begin{displaymath}
I_M = \int_{|T_w{_n}| \ge 1.0} |J_z| dS,
\end{displaymath}
is also derived in the same manner as the magnetic flux $\Psi_M$,  where $J_z$ is the normal electric current density calculated from the horizontal magnetic field at the bottom surface.

\end{itemize}

% ---------------------------------------------------------------
% Figure 1
% ---------------------------------------------------------------

% ---------------------------------------------------------------
% Figure 2
% ---------------------------------------------------------------
\begin{figure}[ht!]
\plotone{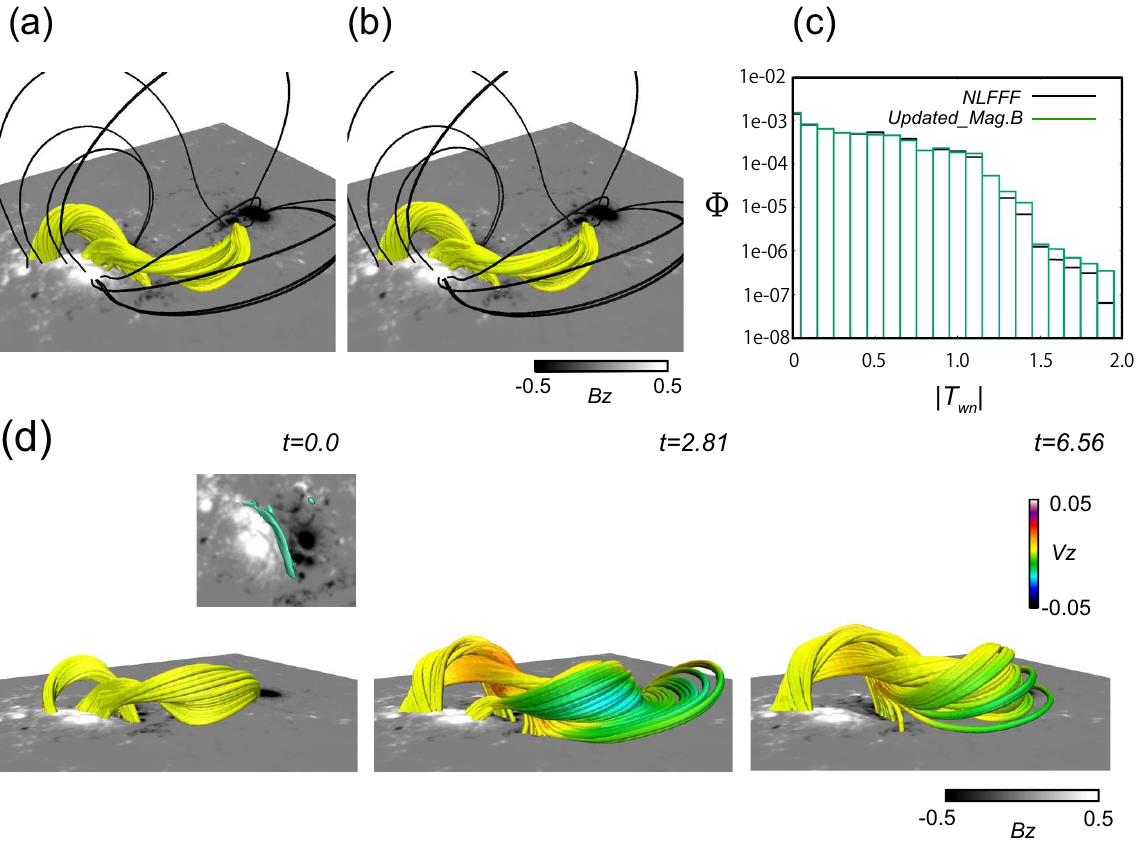}
\caption{(a)-(b) 3D magnetic field structures of the NLFFF and the updated magnetic field from the data-driven MHD simulation. Yellow lines show sheared field lines, and black lines represent overlying field lines. (c) Histograms of magnetic flux $\Phi$ as a function of negative twist magnitude $T_{wn}$ for the NLFFF and the updated field. (d) Temporal evolution of the 3D magnetic field in the pre-eruption phase for the case without small-scale emerging flux (Run R). Field line colors indicate vertical velocity $V_z$. The inset shows a green iso-surface of strong current density ($|{\bf J}| = 20$), concentrated along the polarity inversion line of the eastern sunspots.
\label{fig:f2}}
\end{figure}

% ----------------------------------------------------------------
% Figure 3
% ---------------------------------------------------------------
\begin{figure}[ht!]
\plotone{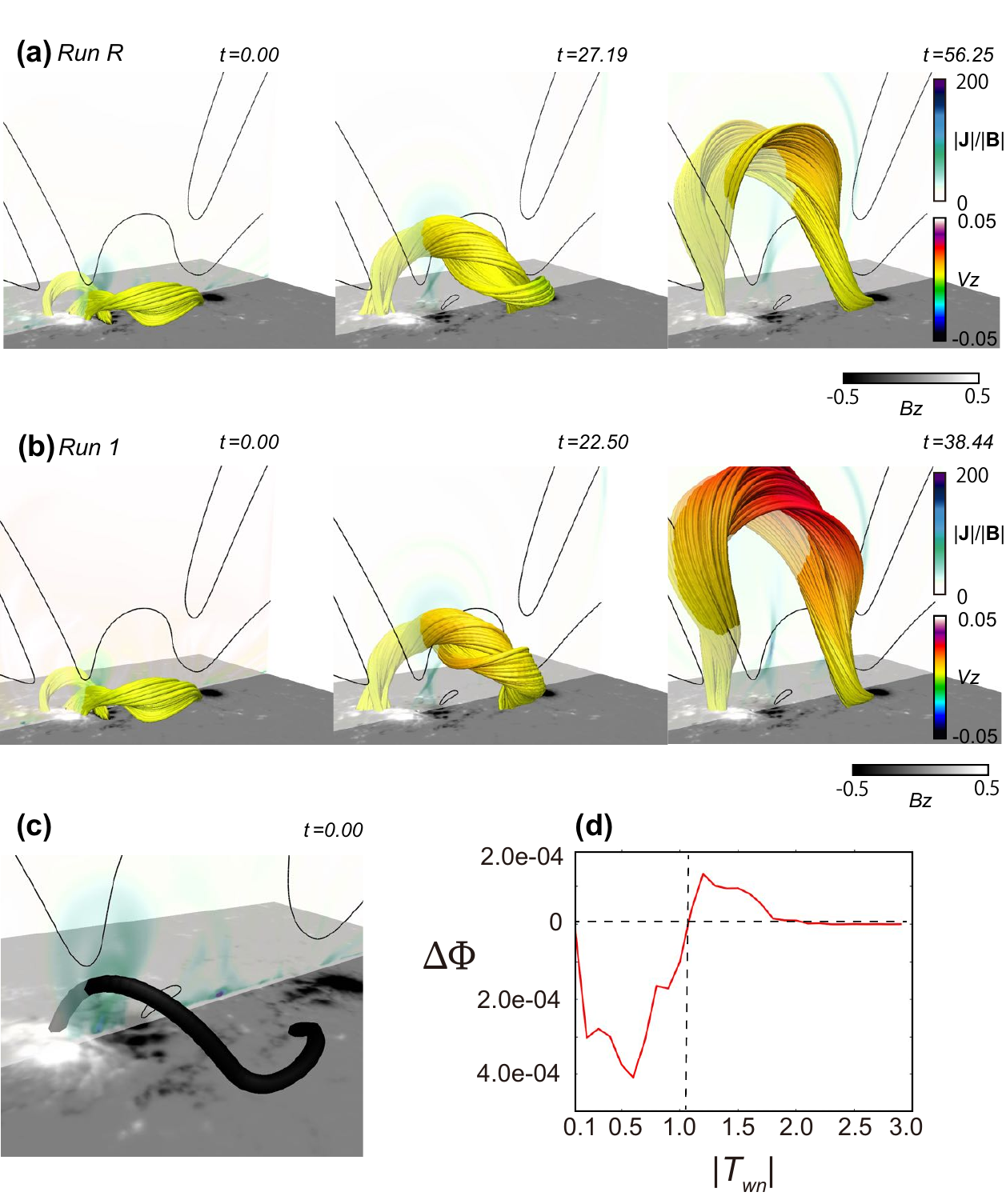}
\caption{(a)-(b) Temporal evolution of 3D erupting magnetic field lines for Runs R and 1. The same color convention as in Figure~\ref{fig:f2}(d) is used to identify the erupting magnetic flux rope. Black contours indicate the decay index of 1.5, and vertical cross-sections show $|{\bf J}|/|{\bf B}|$. Sampling times differ between runs. (c) Part of the initial magnetic field that forms the MFR, traced via Lagrangian tracking of a plasma element initially located at $(x, y, z) = (0.36172, 0.49482, 0.054)$ at $t = 0$. (d) Difference in magnetic flux $\Delta \Phi$ between $t = 0.0$ and $t = 34.96$, plotted against the magnitude of negative twist $T_{wn}$. The horizontal dashed line marks $\Delta \Phi = 0$; its intersection with the vertical line indicates where $\Delta \Phi$ changes sign.\label{fig:f3}}
\end{figure}

% ----------------------------------------------------------------
% Figure 4
% ---------------------------------------------------------------
\begin{figure}[ht!]
\plotone{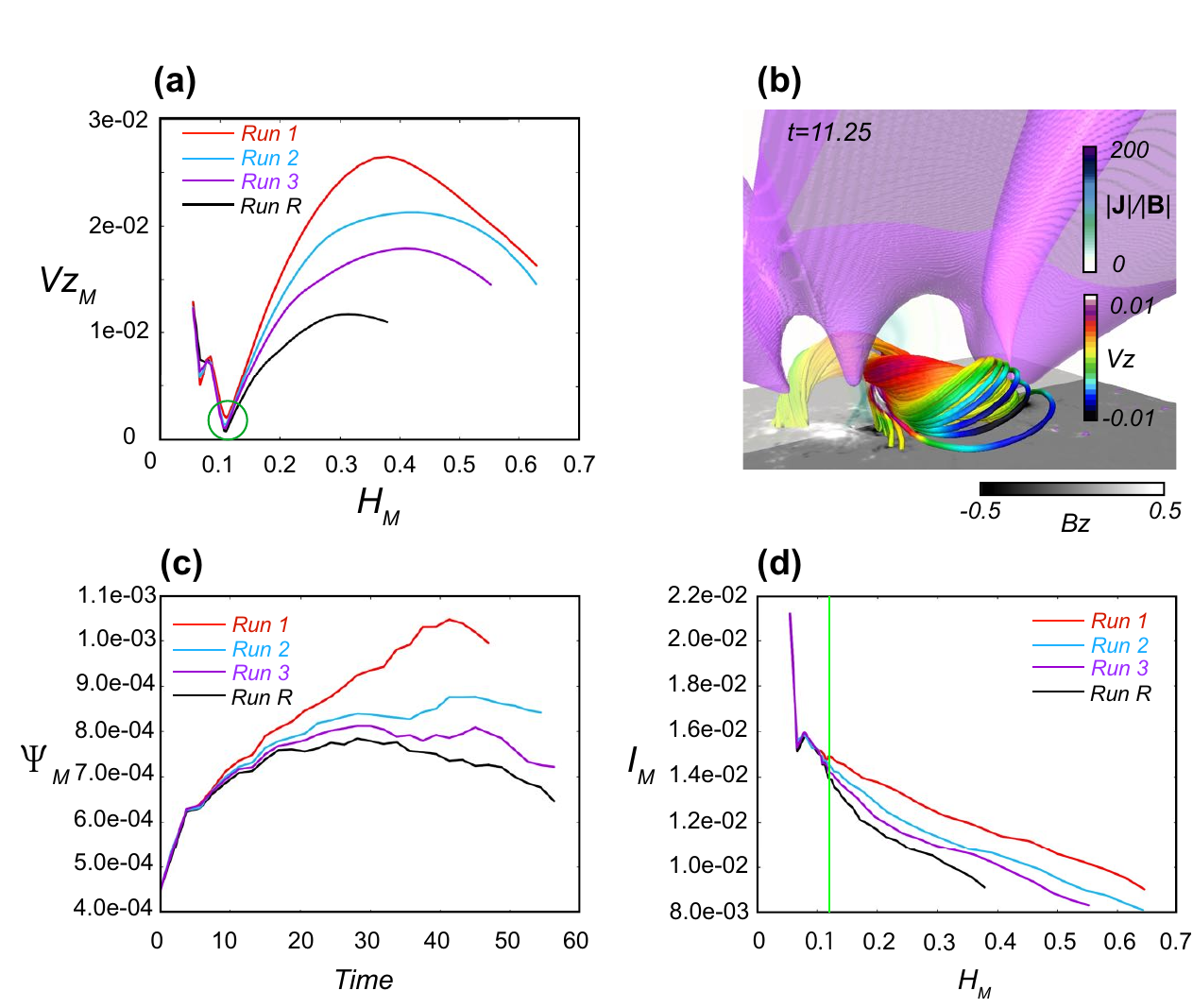}
\caption{(a) MFR velocity profile as a function of height for each run, based on Lagrangian tracking of a plasma element initially located at $(x,y,z)=(0.36172, 0.49482, 0.054)$ at $t=0$ on the thick black line (See Figure \ref{fig:f3}(c)). The red, light blue, purple, and black lines represent Runs 1-3, and R, respectively. (b) Three-dimensional magnetic field structure at $t=11.25$, using the same color coding as in Figure~\ref{fig:f2}(d), when the tracking point passes through at $h \approx 1.1$. The purple surface represents the iso-surface where the decay index equals $1.5$. (c) Temporal evolution of magnetic flux ($\Psi_M$) of the MFR associated with the highly negative twisted field lines satisfying $|T_w{_n}| \ge 1$ with color cording identical to (a). (d) Toroidal current ($I_M$) perpendicular to the MFR cross-section, plotted as a function of height based on the same Lagrangian plasma element. The vertical green line marks the location where the MFR becomes unstable to the torus instability, as discussed in (a). \label{fig:f4}}
\end{figure}

% =================================================
  \section{Results} \label{sec:results}
% =================================================
\subsection{Properties of the Initial Condition}
We briefly discuss the properties of the initial condition used in the data-constrained MHD simulation. Figures~\ref{fig:f2}(a) and (b) show the 3D magnetic structures of the NLFFF and the updated magnetic field, which was evolved for only 0.375 Alfv\'en times in the data-driven MHD simulation. Due to the short driven time, both structures remain nearly identical. A more quantitative comparison is presented in Figure~\ref{fig:f2}(c), which shows a histogram of magnetic flux ($\Phi$) as a function of negative twist ($T_{wn}$) for both the NLFFF and the updated field. The histogram reveals that only the flux associated with strongly negatively twisted field lines ($|T_{wn}| \geq 1.0$) shows a slight increase in the updated field, but the overall flux remains small. We computed the total flux of such highly twisted field lines and found it increased by only a factor of 1.07 after the data-driven simulation. This indicates that the data-driven MHD simulation introduced minimal structural changes. However, since the NLFFF is in---or close to---a force-free equilibrium, such perturbations are necessary to transition the system into a dynamic state.

The initial condition (Figure \ref{fig:f2}(b)) of the data-constrained MHD simulation, taken from the output of the data-driven MHD simulation, is not in equilibrium. Furthermore, the inclusion of a finite resistivity $\eta$ in the induction equation (Eq.\ref{induc_eq}) triggers tether-cutting reconnection \citep{Moore2001} at regions of strong current density located between two systems of sheared magnetic field lines, as shown in the inset of Figure \ref{fig:f2}(d). Figure \ref{fig:f2}(d) shows the 3D magnetic field evolution at an early stage, in which no small-scale emerging flux is introduced, corresponding to Run R. Even in this case, highly twisted field lines begin to form. Thus, in our main simulation, we introduce small-scale emerging flux into a pre-existing magnetic configuration that is already in a dynamic, non-equilibrium state during the pre-eruption phase. This distinguishes our study from most previous works, which primarily focused on the eruption onset. Here, we investigate for the first time the role of small-scale flux emergence in the evolution of an already destabilizing MFR.

\subsection{Three-dimensional Dynamics of Erupting Magnetic Flux Rope}
Figures~\ref{fig:f3}(a)$-$(b) show representative results from the data-constrained MHD simulations for Runs R and 1. The colored lines indicate magnetic field lines, with color representing plasma velocity, and the vertical slice shows $|{\bf J}|/|{\bf B}|$. In both runs, highly twisted field lines are formed via tether-cutting reconnection. As they rise, strong current densities develop around and beneath them. When these structures reach parts of the black contour, the current distribution beneath the twisted lines transitions from a cross-shaped structure to a vertically elongated current sheet.

This morphological change is accompanied---especially in Run 1---by a significant increase in the upward velocity of the twisted field lines. One possible mechanism for this acceleration is the torus instability \citep{Kliem2006}, which occurs when the outward hoop force on the MFR exceeds the restraining strapping force from the overlying field. The black contour in the figure marks where the decay index $n(x, y = 0.355, z) = 1.5$, the theoretical threshold for instability. Since the core of the twisted field lines reaches this contour, they may become torus-unstable. However, this estimate is approximate, and a more quantitative analysis is needed to clarify the onset and role of the instability.

Although minor differences appeared during the early evolution, the overall process was similar with and without the small-scale emerging flux (see animation). However, a clear difference emerged during the main eruption phase (right column). In Run R (without emerging flux), the twisted field lines rose and slowly expanded after crossing the black line. In contrast, in Runs 1 (with emerging flux), the field lines underwent a rapid eruption. These results suggest that the small-scale emerging flux played a key role in triggering the rapid eruption, despite identical initial conditions and decay index distributions.

% -------------------------------------------------------------------
\subsection{Quantitative Evolution of the Magnetic Flux Rope}
% -------------------------------------------------------------------
We traced the height and velocity of the MFR, and quantified its evolution by calculating the magnetic flux ($\Psi_M$) and toroidal current intensity ($I_M$) within the MFR. Here, $\Psi_M$ and $I_M$ refer to the flux and current intensity confined to the region occupied by strongly negatively twisted field lines. Note that $\Psi_M$ differs from the total magnetic flux $\Phi$, which includes a broader area encompassing all negatively twisted field lines.

To trace the height and velocity of MFR, we used Lagrangian tracking of the plasma element that was initially located at a specific point of the black line at $t=0$, which is part of the MFR shown in Figure \ref{fig:f3}(c). Furthermore, 
to find $\Psi_M$ and $I_M$, we need to define the MFR. We computed the difference between histograms of magnetic flux $\Phi$ versus $|T_{wn}|$ at $t = 0.0$ and $t = 34.96$. Figure~\ref{fig:f3}(d) shows the result. The boundary between increased and decreased magnetic flux appears around $|T_{wn}| = 1.0$, suggesting that field lines with $|T_{wn}| \gtrsim 1.0$ contributed to the erupting MFR. We therefore defined the MFR as the bundle of twisted field lines satisfying $|T_{wn}| \ge 1.0$.

Figure~\ref{fig:f4}(a) shows the velocity profiles for each run as a function of MFR height. The height at which the MFRs began to accelerate was nearly identical, around $h \approx 0.11$ (green circles), indicating that the onset of instability occurred at roughly the same location in all runs.
Notably, all runs exhibit nearly identical behavior during the early phase, with only minimal differences that are barely discernible. However, after the MFRs reach a height of approximately $h \approx 1.1$, their accelerations begin to diverge. Figure~\ref{fig:f4}(b) shows a snapshot of the 3D MFR in Run 1 when the tracked point passed through $h \approx 0.11$. At this time, the MFR center was approaching the iso-surface where the decay index equals 1.5, and its velocity was increasing, suggesting that the acceleration was driven by torus instability. Although the critical threshold for torus instability is known to depend strongly on the decay index distribution \citep{Kliem2006, Torok2007}, our results show that the instability growth rate was enhanced by the small-scale emerging flux---even under identical initial conditions and decay index distributions.

The temporal evolution of the MFR magnetic flux $\Psi_M$ is shown in Figure~\ref{fig:f4}(c). As defined above, $\Psi_M$ represents the flux of field lines with $|T_{wn}| \ge 1.0$. The early evolution (up to $t \approx 10$) was similar across all runs, but differences gradually emerged by $t \approx 20$, and became more pronounced afterward due to variations in eruption speed and resulting reconnection rates. Run 1 produced the most enhanced MFR with strongly twisted field lines, while Run R showed the weakest enhancement, leading to different eruption speeds.

Figure~\ref{fig:f4}(d) shows the toroidal current $I_M$ (perpendicular to the MFR cross-section) as a function of MFR height. The green line marks the onset height of the torus instability, as determined from Figure~\ref{fig:f4}(a). At this point, $I_M$ differed slightly among the runs. Since the toroidal current contributes to the hoop force that drives the torus instability \citep{Kliem2006,Welsch2018, ZhongZe2023}, even this small difference in $I_M$ can significantly affect the eruption dynamics. These results suggest that although the small-scale emerging flux had only a minor impact during the pre-eruption phase, it critically influenced the subsequent torus instability by modulating the toroidal current.

% ===============================================================
\section{Discussion and Conclusion}
% ===============================================================
% ----------------------------------------------------------------
% Figure 5
% ---------------------------------------------------------------
\begin{figure}[ht!]
\plotone{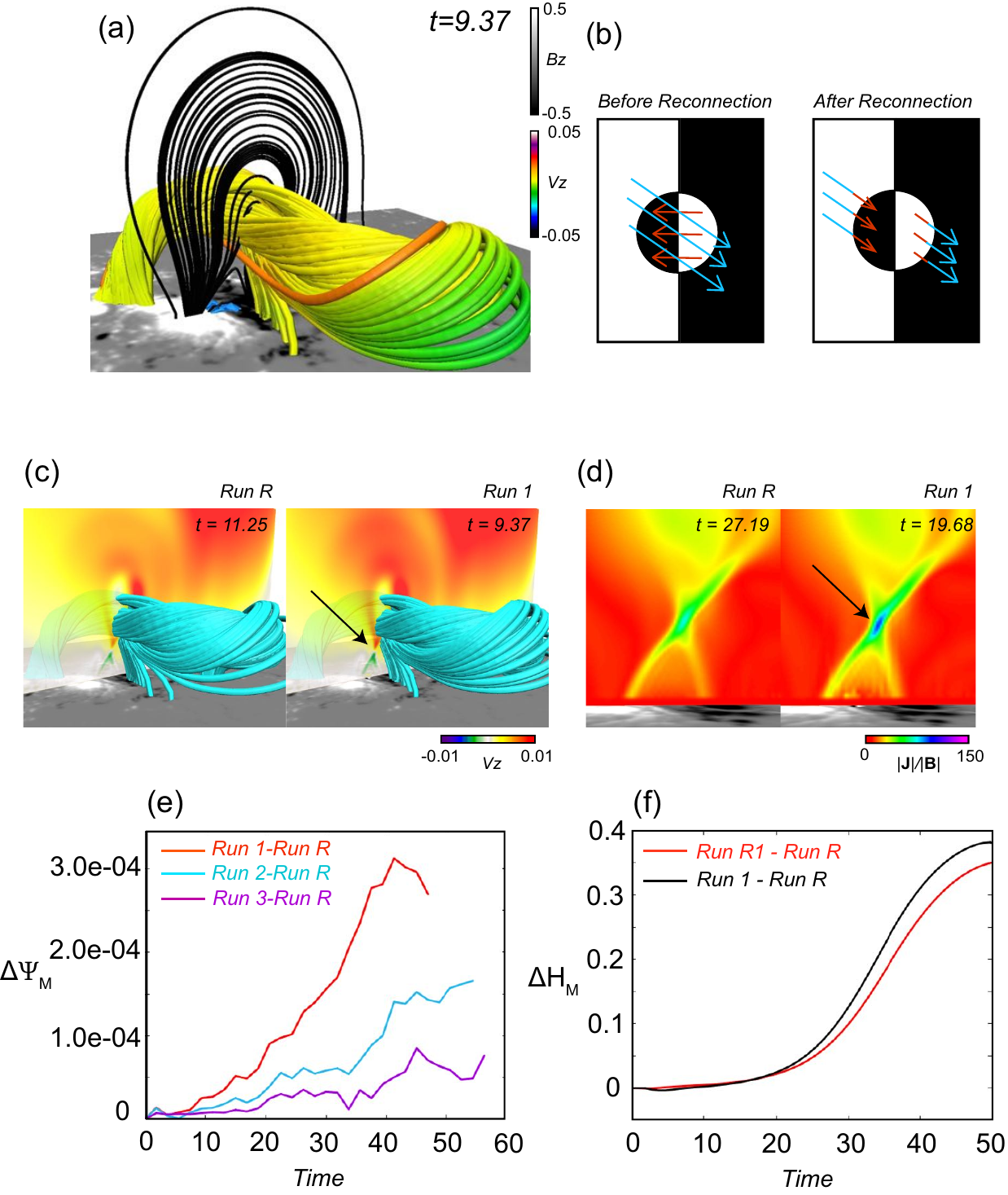}
\caption{(a) 3D magnetic field structure at $t=9.37$ in the pre-eruption phase. The MFR is colored by $V_z$; the orange line indicates a field line just after tether-cutting reconnection. Black lines show surrounding field lines and post-flare loops (projected on the $x-z$ plane), while blue lines are traced the small-scale emerging flux. An enlarged view is shown in Figure~\ref{fig:f1}. (b) Schematic of reconnection between the small-scale emerging flux and pre-existing coronal loops. White/black regions denote positive/negative polarity; red and blue arrows indicate the emerging and coronal fields, respectively. (c) 3D field lines in sky blue for Runs R ($t=11.25$) and 1 ($t=9.37$), showing MFR formation via tether-cutting reconnection. Vertical cross-sections show $V_z$; black arrow marks reconnection sites with bi-directional flows for Run 1.(d) $|{\bf J}|/|{\bf B}|$ distributions for Run R ($t=27.19$) and Run 1 ($t=19.68$) on vertical slices through the emerging flux region. Black arrow indicates the enhanced region of magnetic reconnection for Run 1. (e) The temporal evolution of the difference in the magnetic flux of the MFR ($\Psi_M$) for each run relative to Run R. The red, sky blue, and purple lines indicate the differences in $\Psi_M$ between Run 1 and Run R, Run 2 and Run R, and Run 3 and Run R, respectively. (f) The temporal evolution of the difference in the height of the MFR between Run 1 and Run R in black, and between Run R1 and R in red.
\label{fig:f5}}
\end{figure}

% ---------------------------------------------------------------------------
As shown in the preceding section, although all runs exhibit similar evolution during the pre-eruption phase, slight differences during this stage play a critical role in producing the significant divergence observed in the subsequent major eruption of the MFR. The origin of these subtle differences is discussed in this section.

Figure~\ref{fig:f5}(a) shows the magnetic structure in Run1 during the pre-eruption phase. The thick orange line traces a field line immediately after tether-cutting reconnection, forming a helical structure that encloses and reinforces the MFR. As shown in Figures \ref{fig:f4}(c) and (d), this reconnection generates twisted field lines and enhances the toroidal current of the MFR. Thus, stronger tether-cutting reconnection contributes to the formation of an MFR with a more intense axial current prior to the onset of torus instability. Therefore, the key question becomes why the small-scale emerging flux leads to a slight enhancement of the tether-cutting reconnection.

The blue lines in Figure~\ref{fig:f5}(a) trace magnetic field lines from the small-scale emerging flux, with an enlarged view shown in Figure~\ref{fig:f1}. Figure \ref{fig:f1}(c) shows, initially forming a closed-loop structure, these field lines open up rapidly ($t \le 1.0$) due to reconnection with pre-existing coronal field lines. This occurs because the polarity of the small-scale emerging flux is opposite to that of the main field, making reconnection favorable \citep{Kusano2012}, as illustrated in Figure~\ref{fig:f5}(b). As a result, as shown in Figure \ref{fig:f5}(c), reconnection-driven bi-directional flows in Run 1 are slightly enhanced compared to Run R, as shown by the $V_z$ distribution on the vertical cross section. Furthermore, as shown in Figure \ref{fig:f5}(d), in the late pre-eruption phase, the current density beneath the MFR is marginally higher in Run 1.

It is notable that the field lines in the small-scale emerging flux region transition from closed to open. This situation may be similar to the one proposed by \cite{Chen2000}, in which small-scale emerging flux enhances reconnection beneath the MFR, thereby triggering its eruption. The field lines of the emerging flux subsequently transition from closed to open (see Figure 3 in their paper). The following scenario can be considered as a possible explanation.
The emerging flux reconnects with the pre-existing magnetic field, which removes the overlying or adjacent magnetic flux near the emergence site. This reduction in magnetic pressure facilitates the tether-cutting reconnection.

In other words, a larger amount of emerging flux can remove more of the surrounding magnetic flux, resulting in a stronger enhancement of the reconnection process. Figure~\ref{fig:f5}(e) shows the temporal evolution of the difference in the magnetic flux of the MFR, $\Delta \Psi_M$, for each run relative to Run R, in which no emerging flux was introduced. Although the initial increase in poloidal flux relative to Run R is modest in each case, the difference gradually grows over time and the rate of increase depends on the magnitude of the imposed emerging flux. We suggest that this difference is progressively amplified through the feedback process of tether-cutting reconnection, leading to increasing differences in both the poloidal magnetic flux and the toroidal current of the MFR. These differences ultimately play a critical role in the growth rate of the torus instability.

Finally, we discuss the effect of the slightly enhanced tether-cutting reconnection. To do this, we conducted an additional experiment. Specifically, we introduced a small velocity perturbation, defined as
\begin{displaymath}
v_z^e = -1.0\times 10^{-2} exp \left[{\frac{-\left\{(x-x_0)^2+(y-y_0)^2+(z-z_0)^2\right\}}{2.5\times 10^{-5}}}\right]exp\left\{-\frac{t-1.0}{10.0}\right\},
\end{displaymath}
into Run R. This perturbation was applied with the spatial peak center at $(x_0, y_0, z_0) = (x_e, y_e, 1.0 \times 10^{-2})$ and the temporal peak at $t = 1.0$, corresponding to the location just above the emerging flux region in Runs 1$-$3. The precise position is located at a height equal to twice the radius of the emerging flux from the solar surface. Although it reaches its maximum magnitude ($1.0\times 10^{-2}$) at the beginning, the value decreases exponentially over time. We refer to this experiment as Run R1. This perturbation enhances the downflow associated with tether-cutting reconnection, thereby intensifying the reconnection itself. Therefore, in this experiment, we investigate both the initial impact of the tether-cutting reconnection and the effect of its sustained action.

Figure~\ref{fig:f5}(f) shows the temporal evolution of the difference in the MFR height, $\Delta H_M$, between Run 1 and Run R, as well as between Run R1 and Run R. The striking similarity in the behavior of both cases implies that the emerging flux acts in a manner similar to the perturbation applied in Run R1. No significant differences in $\Delta H_M$ were observed during the early phase in either case, indicating that tether-cutting reconnection was not yet prominent. However, in both cases, the difference from Run R gradually increased and exhibited a sudden jump at $t \approx 25$. This result suggests that even a slight and continuous enhancement of tether-cutting reconnection may play an important role. The degree of enhancement might depend on how much the initially injected magnetic flux has removed the surrounding magnetic flux. In a recent study, a magnetic field strength of up to 6000G was observed locally in an active region \citep{Okamoto2018}. If such a strong field were to emerge in a flare-productive region, fast down flows could form due to reconnection between the pre-existing field and the strongly emerging flux, potentially resulting in faster acceleration than reported in this study. 

The results and interpretations presented here were made possible by the combination of two state-of-the-art approaches: high-resolution observations from BBSO and data-based MHD simulations. In the near future, joint studies incorporating newer ground-based observations, such as those from the Daniel K. Inouye Solar Telescope \citep{DKIST2021}, are expected to lead to significant breakthroughs. Thus, both observational and simulation-based approaches are becoming increasingly essential for advancing our understanding of solar eruptions.

We mention that the numerical resolution limits the precise characterization of reconnection dynamics between small-scale emerging flux and pre-existing magnetic field. However, the core physical mechanism revealed in this study remains robust under these conditions. These results are not intended as a definitive proof, but rather as a physically plausible scenario based on the currently accessible resolution. Further studies employing higher-resolution simulations and different modeling approaches will be important to quantitatively assess the robustness of the proposed mechanism.

%\section{Software and third party data repository citations} \label{sec:cite}

% -----------------------------------------------------------------------
% Parameters for NLFFF  
% -----------------------------------------------------------------------
  \begin{table}
  \begin{center}
  \caption{Parameters for each Run. $|B_{emf}|$, $\Phi_{emf}$ , $r_c$, $V_{emf}$ are magnitude, magnetic flux, radius, upward velocity of the emerging flux, respectively. $v_z^e$ is additional upward velocity as a perturbation. $B^*, \Phi^* L^*$, and $\tau_A^*$ are normalized values and set to $0.28$T, $3.4\times 10^{16}$Wb, $3.48 \times 10^8 $m, and $348$s.
  \label{tbl-1}}
  \begin{tabular}{crrrrrr}
  \tableline\tableline
   Simulation Runs & Emerging flux & $|B_{emf}|$ & $\Phi_{emf}$ &$r_c$ & $V_{emf}$ & $v_{z}^e$ \\
  \tableline
   Run R    & No & $-$ & $-$ & $-$ & $-$ & No  \\
   Run R1   & No & $-$  & $-$ & $-$ & $-$ & Yes \\
   Run 1    & Yes & $0.75B^*$ & $5.2\times 10^{-5}\Phi^*$ & $5.0\times 10^{-3}L^*$ & $5.0\times 10^{-3}\tau_A^*$ &  No \\
   Run 2    & Yes & $0.50B^*$ & $3.5\times 10^{-5}\Phi^*$ & $5.0\times 10^{-3}L^*$ & $5.0 \times 10^{-3}\tau_A^*$ &  No  \\
   Run 3    & Yes & $0.25B^*$ & $1.73\times 10^{-5}\Phi^*$ & $5.0\times 10^{-3}L^*$ & $5.0 \times 10^{-3}\tau_A^*$ & No \\ 
   \tableline
   \end{tabular} 
   \end{center}
   \end{table}

% --------------------------------------------------------------
\begin{acknowledgments}
% --------------------------------------------------------------
 We thank anonymous referee for useful comments.
 This work was supported by the National Science Foundation  under grant AGS-1954737, 2145253, 2149748, 2206424, 2300341, 2408174,  2309939, 2401229, and AST-2204384, 2206424, 2149748, 2108235, and National Aeronautics and Space Administration under grants 80NSSC20K0025, 80NSSC21K1671, 80NSSC21K0003, 80HQTR20T0067, 80NSSC24M0174, 80NSSC23K0406, 80NSSC24K1914. TM is supported by JSPS KAKENHI Grant Numbers JP20K11851, JP20H00156, JP24K07117. We acknowledge the use of data from the Goode Solar Telescope (GST) of the Big Bear Solar Observatory (BBSO). BBSO operation is supported by US NSF AGS-2309939 and New Jersey Institute of Technology. GST operation is partly supported by the Korea Astronomy and Space Science Institute and the Seoul National University. The 3D visualizations were produced using VAPOR (\href{http://www.vapor.ucar.edu}{\texttt{www.vapor.ucar.edu}}), a product of the National Center for Atmospheric Research (\citealt{Li2019}). All numerical calculations in this paper were performed using the computing facilities of the High Performance Computing Center (HPCC) at the New Jersey Institute of Technology.
\end{acknowledgments}


%\vspace{5mm}
%\facilities{HST(STIS), Swift(XRT and UVOT), AAVSO, CTIO:1.3m,


%%%\software{astropy \citep{2013A&A...558A..33A,2018AJ....156..123A},  
%%%          Cloudy \citep{2013RMxAA..49..137F}, 
%%%          Source Extractor \citep{1996A&AS..117..393B}
%%%          }

%\bibliography{sample631}{}
\bibliographystyle{aasjournal}
\bibliography{sample631}

\begin{thebibliography}{}
\expandafter\ifx\csname natexlab\endcsname\relax\def\natexlab#1{#1}\fi
\providecommand{\url}[1]{\href{#1}{#1}}
\providecommand{\dodoi}[1]{doi:~\href{http://doi.org/#1}{\nolinkurl{#1}}}
\providecommand{\doeprint}[1]{\href{http://ascl.net/#1}{\nolinkurl{http://ascl.net/#1}}}
\providecommand{\doarXiv}[1]{\href{https://arxiv.org/abs/#1}{\nolinkurl{https://arxiv.org/abs/#1}}}

\bibitem[{{Ahn} {et~al.}(2016){Ahn}, {Cao}, {Shumko}, \& {Chae}}]{Anh2016}
{Ahn}, K., {Cao}, W., {Shumko}, S., \& {Chae}, J. 2016, in AAS/Solar Physics Division Meeting, Vol.~47, AAS/Solar Physics Division Abstracts \#47, 2.07

\bibitem[{{Amari} {et~al.}(1996){Amari}, {Luciani}, {Aly}, \& {Tagger}}]{Amari1996}
{Amari}, T., {Luciani}, J.~F., {Aly}, J.~J., \& {Tagger}, M. 1996, \apjl, 466, L39, \dodoi{10.1086/310158}

\bibitem[{{Aulanier} {et~al.}(2010){Aulanier}, {T{\"o}r{\"o}k}, {D{\'e}moulin}, \& {DeLuca}}]{Aulanier2010}
{Aulanier}, G., {T{\"o}r{\"o}k}, T., {D{\'e}moulin}, P., \& {DeLuca}, E.~E. 2010, \apj, 708, 314, \dodoi{10.1088/0004-637X/708/1/314}

\bibitem[{{Bamba} \& {Kusano}(2018)}]{Bamba2018}
{Bamba}, Y., \& {Kusano}, K. 2018, \apj, 856, 43, \dodoi{10.3847/1538-4357/aaacd1}

\bibitem[{{Berger} \& {Prior}(2006)}]{Berger2006}
{Berger}, M.~A., \& {Prior}, C. 2006, Journal of Physics A Mathematical General, 39, 8321, \dodoi{10.1088/0305-4470/39/26/005}

\bibitem[{{Bobra} {et~al.}(2014){Bobra}, {Sun}, {Hoeksema}, {Turmon}, {Liu}, {Hayashi}, {Barnes}, \& {Leka}}]{Bobra2014}
{Bobra}, M.~G., {Sun}, X., {Hoeksema}, J.~T., {et~al.} 2014, \solphys, 289, 3549, \dodoi{10.1007/s11207-014-0529-3}

\bibitem[{{Cao} {et~al.}(2012){Cao}, {Goode}, {Ahn}, {Gorceix}, {Schmidt}, \& {Lin}}]{Cao2012}
{Cao}, W., {Goode}, P.~R., {Ahn}, K., {et~al.} 2012, in Astronomical Society of the Pacific Conference Series, Vol. 463, Second ATST-EAST Meeting: Magnetic Fields from the Photosphere to the Corona., ed. T.~R. {Rimmele}, A.~{Tritschler}, F.~{W{\"o}ger}, M.~{Collados Vera}, H.~{Socas-Navarro}, R.~{Schlichenmaier}, M.~{Carlsson}, T.~{Berger}, A.~{Cadavid}, P.~R. {Gilbert}, P.~R. {Goode}, \& M.~{Kn{\"o}lker}, 291

\bibitem[{{Cao} {et~al.}(2010){Cao}, {Gorceix}, {Coulter}, {Ahn}, {Rimmele}, \& {Goode}}]{Cao2010}
{Cao}, W., {Gorceix}, N., {Coulter}, R., {et~al.} 2010, Astronomische Nachrichten, 331, 636, \dodoi{10.1002/asna.201011390}

\bibitem[{{Chen} \& {Shibata}(2000)}]{Chen2000}
{Chen}, P.~F., \& {Shibata}, K. 2000, \apj, 545, 524, \dodoi{10.1086/317803}

\bibitem[{{Dedner} {et~al.}(2002){Dedner}, {Kemm}, {Kr{\"o}ner}, {Munz}, {Schnitzer}, \& {Wesenberg}}]{Dedner2002}
{Dedner}, A., {Kemm}, F., {Kr{\"o}ner}, D., {et~al.} 2002, Journal of Computational Physics, 175, 645, \dodoi{10.1006/jcph.2001.6961}

\bibitem[{{Fan}(2010)}]{Fan2010}
{Fan}, Y. 2010, \apj, 719, 728, \dodoi{10.1088/0004-637X/719/1/728}

\bibitem[{{Goode} {et~al.}(2010){Goode}, {Coulter}, {Gorceix}, {Yurchyshyn}, \& {Cao}}]{Goode2010}
{Goode}, P.~R., {Coulter}, R., {Gorceix}, N., {Yurchyshyn}, V., \& {Cao}, W. 2010, Astronomische Nachrichten, 331, 620, \dodoi{10.1002/asna.201011387}

\bibitem[{{Hayashi} {et~al.}(2018){Hayashi}, {Feng}, {Xiong}, \& {Jiang}}]{Hayashi2018}
{Hayashi}, K., {Feng}, X., {Xiong}, M., \& {Jiang}, C. 2018, \apj, 855, 11, \dodoi{10.3847/1538-4357/aaacd8}

\bibitem[{{Inoue}(2016)}]{Inoue2016}
{Inoue}, S. 2016, Progress in Earth and Planetary Science, 3, 19, \dodoi{10.1186/s40645-016-0084-7}

\bibitem[{{Inoue} {et~al.}(2014){Inoue}, {Hayashi}, {Magara}, {Choe}, \& {Park}}]{Inoue2014}
{Inoue}, S., {Hayashi}, K., {Magara}, T., {Choe}, G.~S., \& {Park}, Y.~D. 2014, \apj, 788, 182, \dodoi{10.1088/0004-637X/788/2/182}

\bibitem[{{Inoue} {et~al.}(2023){Inoue}, {Hayashi}, {Miyoshi}, {Jing}, \& {Wang}}]{Inoue2023b}
{Inoue}, S., {Hayashi}, K., {Miyoshi}, T., {Jing}, J., \& {Wang}, H. 2023, \apjl, 944, L44, \dodoi{10.3847/2041-8213/acb7f4}

\bibitem[{{Inoue} {et~al.}(2018){Inoue}, {Kusano}, {B{\"u}chner}, \& {Sk{\'a}la}}]{Inoue2018}
{Inoue}, S., {Kusano}, K., {B{\"u}chner}, J., \& {Sk{\'a}la}, J. 2018, Nature Communications, 9, 174, \dodoi{10.1038/s41467-017-02616-8}

\bibitem[{{Jing} {et~al.}(2016){Jing}, {Xu}, {Cao}, {Liu}, {Gary}, \& {Wang}}]{Jing2016}
{Jing}, J., {Xu}, Y., {Cao}, W., {et~al.} 2016, Scientific Reports, 6, 24319, \dodoi{10.1038/srep24319}

\bibitem[{{Jing} {et~al.}(2021){Jing}, {Inoue}, {Lee}, {Li}, {Nita}, {Xu}, {Liu}, {Gary}, \& {Wang}}]{Jing2021}
{Jing}, J., {Inoue}, S., {Lee}, J., {et~al.} 2021, \apj, 922, 108, \dodoi{10.3847/1538-4357/ac26c7}

\bibitem[{{Kang} {et~al.}(2019){Kang}, {Inoue}, {Kusano}, {Park}, \& {Moon}}]{Kang2019}
{Kang}, J., {Inoue}, S., {Kusano}, K., {Park}, S.-H., \& {Moon}, Y.-J. 2019, \apj, 887, 263, \dodoi{10.3847/1538-4357/ab5582}

\bibitem[{{Kliem} \& {T{\"o}r{\"o}k}(2006)}]{Kliem2006}
{Kliem}, B., \& {T{\"o}r{\"o}k}, T. 2006, \prl, 96, 255002, \dodoi{10.1103/PhysRevLett.96.255002}

\bibitem[{{Kubo} {et~al.}(2007){Kubo}, {Yokoyama}, {Katsukawa}, {Lites}, {Tsuneta}, {Suematsu}, {Ichimoto}, {Shimizu}, {Nagata}, {Tarbell}, {Shine}, {Title}, \& {Elmore David}}]{Kubo2007}
{Kubo}, M., {Yokoyama}, T., {Katsukawa}, Y., {et~al.} 2007, \pasj, 59, S779, \dodoi{10.1093/pasj/59.sp3.S779}

\bibitem[{{Kusano} {et~al.}(2012){Kusano}, {Bamba}, {Yamamoto}, {Iida}, {Toriumi}, \& {Asai}}]{Kusano2012}
{Kusano}, K., {Bamba}, Y., {Yamamoto}, T.~T., {et~al.} 2012, \apj, 760, 31, \dodoi{10.1088/0004-637X/760/1/31}

\bibitem[{{Li} {et~al.}(2019){Li}, {Jaroszynski}, {Pearse}, {Orf}, \& {Clyne}}]{Li2019}
{Li}, S., {Jaroszynski}, S., {Pearse}, S., {Orf}, L., \& {Clyne}, J. 2019, Atmosphere, 10, 488, \dodoi{10.3390/atmos10090488}

\bibitem[{{Liu} {et~al.}(2022){Liu}, {Jing}, {Xu}, \& {Wang}}]{Liu.Nian2022}
{Liu}, N., {Jing}, J., {Xu}, Y., \& {Wang}, H. 2022, \apj, 930, 154, \dodoi{10.3847/1538-4357/ac6425}

\bibitem[{{Moore} {et~al.}(2001){Moore}, {Sterling}, {Hudson}, \& {Lemen}}]{Moore2001}
{Moore}, R.~L., {Sterling}, A.~C., {Hudson}, H.~S., \& {Lemen}, J.~R. 2001, \apj, 552, 833, \dodoi{10.1086/320559}

\bibitem[{{Muhamad} {et~al.}(2017){Muhamad}, {Kusano}, {Inoue}, \& {Shiota}}]{Muhamad2017}
{Muhamad}, J., {Kusano}, K., {Inoue}, S., \& {Shiota}, D. 2017, \apj, 842, 86, \dodoi{10.3847/1538-4357/aa750e}

\bibitem[{{Okamoto} \& {Sakurai}(2018)}]{Okamoto2018}
{Okamoto}, T.~J., \& {Sakurai}, T. 2018, \apjl, 852, L16, \dodoi{10.3847/2041-8213/aaa3d8}

\bibitem[{{Pesnell} {et~al.}(2012){Pesnell}, {Thompson}, \& {Chamberlin}}]{Pesnell2012}
{Pesnell}, W.~D., {Thompson}, B.~J., \& {Chamberlin}, P.~C. 2012, \solphys, 275, 3, \dodoi{10.1007/s11207-011-9841-3}

\bibitem[{{Rast} {et~al.}(2021){Rast}, {Bello Gonz{\'a}lez}, {Bellot Rubio}, {Cao}, {Cauzzi}, {Deluca}, {de Pontieu}, {Fletcher}, {Gibson}, {Judge}, {Katsukawa}, {Kazachenko}, {Khomenko}, {Landi}, {Mart{\'\i}nez Pillet}, {Petrie}, {Qiu}, {Rachmeler}, {Rempel}, {Schmidt}, {Scullion}, {Sun}, {Welsch}, {Andretta}, {Antolin}, {Ayres}, {Balasubramaniam}, {Ballai}, {Berger}, {Bradshaw}, {Campbell}, {Carlsson}, {Casini}, {Centeno}, {Cranmer}, {Criscuoli}, {Deforest}, {Deng}, {Erd{\'e}lyi}, {Fedun}, {Fischer}, {Gonz{\'a}lez Manrique}, {Hahn}, {Harra}, {Henriques}, {Hurlburt}, {Jaeggli}, {Jafarzadeh}, {Jain}, {Jefferies}, {Keys}, {Kowalski}, {Kuckein}, {Kuhn}, {Kuridze}, {Liu}, {Liu}, {Longcope}, {Mathioudakis}, {McAteer}, {McIntosh}, {McKenzie}, {Miralles}, {Morton}, {Muglach}, {Nelson}, {Panesar}, {Parenti}, {Parnell}, {Poduval}, {Reardon}, {Reep}, {Schad}, {Schmit}, {Sharma}, {Socas-Navarro}, {Srivastava}, {Sterling}, {Suematsu}, {Tarr}, {Tiwari}, {Tritschler}, {Verth}, {Vourlidas}, {Wang}, {Wang}, {NSO and DKIST
  Project}, {DKIST Instrument Scientists}, {DKIST Science Working Group}, \& {DKIST Critical Science Plan Community}}]{DKIST2021}
{Rast}, M.~P., {Bello Gonz{\'a}lez}, N., {Bellot Rubio}, L., {et~al.} 2021, \solphys, 296, 70, \dodoi{10.1007/s11207-021-01789-2}

\bibitem[{{Scherrer} {et~al.}(2012){Scherrer}, {Schou}, {Bush}, {Kosovichev}, {Bogart}, {Hoeksema}, {Liu}, {Duvall}, {Zhao}, {Title}, {Schrijver}, {Tarbell}, \& {Tomczyk}}]{Scherrer2012}
{Scherrer}, P.~H., {Schou}, J., {Bush}, R.~I., {et~al.} 2012, \solphys, 275, 207, \dodoi{10.1007/s11207-011-9834-2}

\bibitem[{{Shibata} \& {Magara}(2011)}]{Shibata2011}
{Shibata}, K., \& {Magara}, T. 2011, Living Reviews in Solar Physics, 8, 6, \dodoi{10.12942/lrsp-2011-6}

\bibitem[{{Threlfall} {et~al.}(2018){Threlfall}, {Hood}, \& {Priest}}]{Threlfall2018}
{Threlfall}, J., {Hood}, A.~W., \& {Priest}, E.~R. 2018, \solphys, 293, 98, \dodoi{10.1007/s11207-018-1318-1}

\bibitem[{{T{\"o}r{\"o}k} \& {Kliem}(2007)}]{Torok2007}
{T{\"o}r{\"o}k}, T., \& {Kliem}, B. 2007, Astronomische Nachrichten, 328, 743, \dodoi{10.1002/asna.200710795}

\bibitem[{{T{\"o}r{\"o}k} {et~al.}(2024){T{\"o}r{\"o}k}, {Linton}, {Leake}, {Miki{\'c}}, {Lionello}, {Titov}, \& {Downs}}]{Torok2024}
{T{\"o}r{\"o}k}, T., {Linton}, M.~G., {Leake}, J.~E., {et~al.} 2024, \apj, 962, 149, \dodoi{10.3847/1538-4357/ad1826}

\bibitem[{{Wang} {et~al.}(2017){Wang}, {Liu}, {Ahn}, {Xu}, {Jing}, {Deng}, {Huang}, {Liu}, {Kusano}, {Fleishman}, {Gary}, \& {Cao}}]{Wang2017}
{Wang}, H., {Liu}, C., {Ahn}, K., {et~al.} 2017, Nature Astronomy, 1, 0085, \dodoi{10.1038/s41550-017-0085}

\bibitem[{{Welsch}(2018)}]{Welsch2018}
{Welsch}, B.~T. 2018, \solphys, 293, 113, \dodoi{10.1007/s11207-018-1329-y}

\bibitem[{{Zhong} {et~al.}(2023){Zhong}, {Guo}, {Wiegelmann}, {Ding}, \& {Chen}}]{ZhongZe2023}
{Zhong}, Z., {Guo}, Y., {Wiegelmann}, T., {Ding}, M.~D., \& {Chen}, Y. 2023, \apjl, 947, L2, \dodoi{10.3847/2041-8213/acc6ce}

\end{thebibliography}



\end{document}